# Interplay of Magnetic Exchange Interactions and Ni–S–Ni Bond Angles in Polynuclear Nickel(II) Complexes

Yulia Krupskaya,*[a] Alexey Alfonsov,[a] Anupama Parameswaran,[a] Vladislav Kataev,[a] Rüdiger Klingeler,[a] Gunther Steinfeld,[b,#] Norman Beyer,[b] Mathias Gressenbuch,[b,§] Berthold Kersting,*[b] and Bernd Büchner[a]

The ability of bridging thiophenolate groups (RS$^-$) to transmit magnetic exchange interactions between paramagnetic Ni(II) ions has been examined. Specific attention was paid to complexes with large Ni-SR-Ni angles. For this purpose, dinuclear [Ni$_2$L$^1$($\mu$-OAc)·I$_2$][I$_5$] (**2**) and trinuclear [Ni$_3$L$^2$(OAc)$_2$][BPh$_4$]$_2$ (**3**), where H$_2$L$^1$ and H$_2$L$^2$ represent 24-membered macrocyclic amino-thiophenol ligands, were prepared and fully characterized by IR- and UV-vis spectroscopy, X-ray crystallography, static magnetization $M$ measurements and high-field electron spin resonance (HF-ESR). The dinuclear complex **2** has a central N$_3$Ni$_2$($\mu$-S)$_2$($\mu$-OAc)Ni$_2$N$_3$ core with a mean Ni-S-Ni angle of 92°. The macrocycle L$^2$ supports a trinuclear complex **3**, with distorted octahedral N$_2$O$_2$S$_2$ and N$_2$O$_3$S coordination environments for one central and two terminal Ni(II) ions, respectively. The Ni–S–Ni angles are at 132.8° and 133.5°. We find that the variation of the bond angles has a very strong impact on the magnetic properties of the Ni complexes. In the case of the Ni$_2$-complex, temperature $T$ and magnetic field $B$ dependencies of $M$ reveal a ferromagnetic coupling $J$ = - 29 cm$^{-1}$ between two Ni(II) ions (H = $J$S$_1$S$_2$). HF-ESR measurements yield a negative axial magnetic anisotropy ($D$ < 0) which implies a bistable (easy axis) magnetic ground state. In contrast, for the Ni$_3$-complex we find an appreciable antiferromagnetic coupling $J'$ = 97 cm$^{-1}$ between the Ni(II) ions and a positive axial magnetic anisotropy ($D$ > 0) which implies an easy plane situation.

## Introduction

Over the past decades much interest has arisen in the synthesis and characterization of polynuclear transition-metal thiolate complexes.[1-5] The interest in these compounds is mainly due to their biological relevance as simple model compounds for the active sites of metalloenzymes,[6-9] and their intriguing magnetic and electronic properties.[10-12] Macrocyclic ligands are often employed as supporting ligands as their complexes are more stable than those of their acyclic counterparts. In addition the metal ions are fixed in close proximity and can be arranged in almost any topology which has important implications for the metal-metal interactions[13,14] and the binuclear metal reactivity.[15]

We are interested in the coordination chemistry of the hexaaza-dithiophenolate ligand H$_2$L$^1$ and its various derivatives.[16,17] An attractive feature of this ligand is the fact that it forms mixed-ligand complexes [M$_2$L$^1$($\mu$-L')]$^+$ with a bioctahedral N$_3$M($\mu$-S)$_2$($\mu$-L')MN$_3$ core structure, irrespective of the type of the coligand or metal. This allowed us to derive a magneto-structural correlation between the sign of the exchange parameter $J$ and the M-S-M bond angle.[16] For dinickel(II) complexes [Ni$^{II}_2$L$^1$($\mu$-L')]$^+$, for example, the spins on the Ni(II) ions couple ferromagnetically if the average Ni-S-Ni angles are at ~ 90 ± 5°.[18-21] For smaller or larger angles, the orthogonality of the magnetic orbitals will be cancelled thereby giving rise to superexchange interactions via one of the ligands orbitals only. Since those processes usually give rise to the antiferromagnetic exchange interaction between the metal centers, thus with increasing deviation from the 90° bonding geometry the antiferromagnetic contributing to the total exchange will grow and produce a change of the sign of $J$. Though, since pioneering works of Goodenough, Kanamori and Anderson (GKA-rules, see, e.g. [22]) such interplay is rather well understood, the actual dependence of J on the bond angle can be often significantly affected by specific details of the local coordination and particular bonding features.[23,24]

[a]   Y. Krupskaya, A. Parameswaran, A. Alfonsov, Dr. V. Kataev, Dr. R. Klingeler, Prof. Dr. B. Büchner
      Leibniz Institute for Solid State and Materials Research IFW Dresden
      Dresden, D-01171, Germany
      Fax: (+)+49 351 4659 313
      E-mail: y.krupskaya@ifw-dresden.de
[b]   Prof. Dr. B. Kersting, G. Steinfeld, M. Gressenbuch, N. Beyer
      Institute of Inorganic Chemistry
      University of Leipzig
      Leipzig, D-04103, Germany
      Fax: (+)+49 341 973 6199
      E-mail: b.kersting@uni-leipzig.de
#     Current address: Institut für Anorganische und Analytische Chemie, Universität Freiburg, Albertstr. 21, D-79104 Freiburg, Germany
§     Current address: bubbles and beyond GmbH, 04299 Leipzig, Germany

The example of egde-shared cuprate chain systems which have been extensively studied theoretically and experimentally in order to understand magnetic superexchange in nearly rectangular geometry illustrates how details of the structure can strongly affect the actual magnitude of the magnetic exchange. For example, in $La_{14-x}Ca_xCu_{24}O_{41}$[25] or $Li_2CuO_2$[26], direct exchange dominates the ferromagnetic nearest neighbor interaction whereas the influence of the side groups coupled to ligands in the famous spin-Peierls compound $CuGeO_3$ substantially modifies the GKA-rules yielding an effective antiferromagnetic nearest neighbour coupling in the Cu-O chain[27]. In this context, experimental investigations of newly synthesized materials addressing the relationship between a specific bonding topology and magnetic interactions between metal ions are important for the understanding of their magnetic behavior.

Specifically, magneto-structural correlations are of importance for the targeted assembly of molecular-based magnetic materials.[28-30] In Ni-based metal-organic complexes, experimentally, an antiferromagnetic exchange interaction, for example, is observed in tris($\mu$-thiophenolate)-bridged $Ni^{II}$ complexes, where the average angle is below 80°.[31] A magneto-structural correlation has also been made on a matched pair of O-bridged (phenolato) and S-bridged (thiophenolato) binuclear copper(II) complexes of the type [Cu_2(L-X)(pz)], where L-O and L-S are binucleating phenolate and thiophenolate ligands.[32]

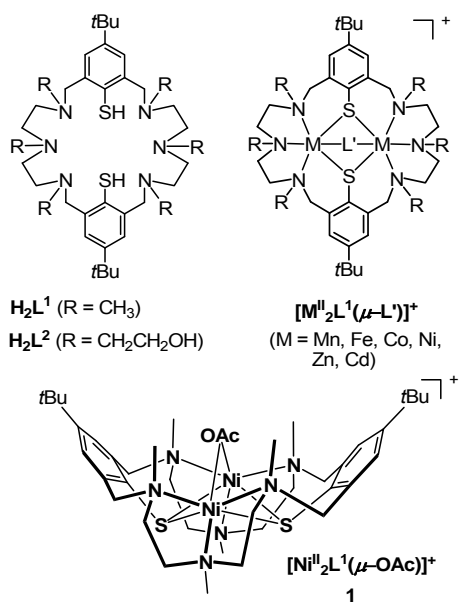

Scheme 1. Structures of Ligands and Complexes.

In view of the above findings it was of interest to synthesize novel nickel(II) amino-thiophenolate complexes with Ni-S-Ni angles appreciably larger than 90° and to characterize their magnetic properties. For $[Ni_2L^1(\mu-L')]^+$ complexes, the most obvious strategy to increase the Ni-S-Ni angles is the use of a large coligand L'.[18] Another strategy to widen the angles could be the attachment of dihalogen molecules to the thiophenolate residues in the form of thiophenolate-dihalogen CT (charge transfer) interactions, as suggested recently by the structure of the $Br_2$ adduct of complex $[Ni_2L^1(\mu-OAc)]^+$ **1**.[33] Thus, upon CT complex formation the average Ni-S-Ni angles increase significantly from 89.6° in **1** to 93.4° in **1**·Br_2. We considered it worthwhile to probe also the utilization of a supporting ligand $H_2L^2$ bearing hydroxyethyl groups in place of the N-methyl functions.[34] Our first attempts afforded a novel iodine adduct $[Ni^{II}_2L^1(\mu-OAc)I_2][I_5]$ (**2**) and a trinuclear complex $[Ni^{II}_3L^2(OAc)_2][BPh_4]_2$ (**3**), both featuring Ni-S-Ni angles wider than 90° as anticipated. Their preparation, spectroscopic properties, and X-ray crystal structures are reported herein. We also present a detailed magnetic study of these complexes by means of static magnetization *M* measurements and high-field/frequency tunable electron spin resonance spectroscopy (HF-ESR). These methods have already been proven as powerful tools for the investigation of magnetic interactions and spin states of molecular Ni(II) complexes.[35-37]

## Results and Discussion

### Synthesis and Characterization of Complexes 2 and 3

The reaction of $[Ni^{II}_2L^1(OAc)][ClO_4]$ (**1**)[38] with five equiv. of diiodine in $CH_3CN$ at 0° C leads to the immediate formation of a dark brown solution, from which black lustrous crystals, characterized as the paramagnetic $I_2$ adduct $[Ni^{II}_2L^1(OAc)·I_2][I_5]$ (**2**), are obtained in 73 % yield (Scheme 2). The trinuclear complex $[Ni^{II}_3L^2(OAc)_2]^{2+}$ formed rather unexpectedly during attempts to prepare a dinuclear complex of the macrocycle $H_2L^2$ (Scheme 1). All complexation attempts using different ligand to metal ratios invariably lead to the formation of the dark-green dication, which could be reproducibly obtained as the tetraphenylborate salt $[Ni^{II}_3L^2(OAc)_2][BPh_4]_2$ (**3**) in yields as high as 60 %.

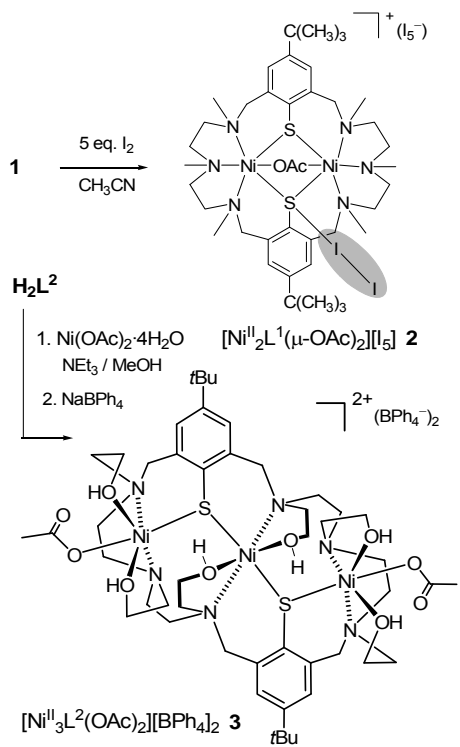

Scheme 2. Synthesis of $[Ni_3L(OAc)_2](BPh_4)_2$ (**1**).

The new complexes **2** and **3** gave satisfactory elemental analyses and were characterized by IR, UV/Vis and ESI-MS spectroscopy and also by X-ray structure analysis. The IR spectrum of **2** shows a strong band at 1579 cm$^{-1}$ attributable to the antisymmetric carboxylate stretching mode.[39] The red-shift of this band (relative to **1**, 1588 cm$^{-1}$)[38] can be traced back to the charge transfer from the thiolate into the antibonding $I_2$ $\sigma^*$ orbital, as the decrease of the charge on the S atom increases the effective charge on the two Ni$^{II}$ ions. This in turn strengthens the Coulomb interactions between the acetate and the nickel ions in **2** and results in the observed red shifts. A similar effect has been observed for the closely related thiolate → $Br_2$ CT adduct of **1**.[33] Complex **3** shows two strong IR bands at 705 and 735 cm$^{-1}$ for the $BPh_4^-$ group and a sharp band at 1579 cm$^{-1}$ attributable to the asymmetric stretching vibration of a coordinated acetate group.[39] The UV/Vis spectrum of **2** in $CH_3CN$ shows a weak band at 1197 nm, which is tentatively assigned to the $\nu_1$ ($^3A_{2g} \rightarrow {}^3T_{2g}(P)$) transition of an octahedral Ni$^{2+}$ ion. The red-shift of this band (relative to **1**, $\nu_1$ = 1120 nm)[38] is taken as an indication that the CT complex **2** remains intact in solution. The bands at 294 nm and 362 nm can be attributed to the spin-allowed $\sigma_g \rightarrow \sigma^*_u$ and $\pi_g \rightarrow \sigma^*_u$ absorptions of the $I_5^-$ ions.[40-42] The UV/vis spectrum of **3** in acetonitrile is marked by three weak absorption bands at 644, 903 and 1073 nm which can be ascribed to the $^3A_{2g} \rightarrow {}^3T_{1g}(F)$ and $^3A_{2g} \rightarrow {}^3T_{2g}(P)$ transitions (the latter split by lower symmetry) of octahedral Ni$^{2+}$ ($3d^8$, $S_{Ni}$ = 1) ions.[43] Values below 500 nm are either associated with $\pi-\pi$ transitions within the ligand or with $RS^- \rightarrow Ni^{2+}$ ligand-to-metal charge transfer transitions.

To ascertain the formulation of compounds **2** and **3**, X-ray crystallographic studies were undertaken. The crystal structure determination of **2** confirmed the presence of the diiodine adduct [Ni$_2$L$^1$($\mu$-OAc)(I$_2$)]$^+$ (Figure 1) and a V-shaped pentaiodide counteranion. One $I_2$ molecule is coordinated to one of the thiolate S atoms with almost perfect linearity of the S–I–I bonds (RS–I–I = 179.54(3)°). The S–I bond distance is 2.755(2) Å, while the I–I distance is lengthened to 2.894(1) Å relative to 2.667(2) Å in free $I_2$ in agreement with the formulation of [Ni$_2$L$^1$($\mu$-OAc)(I$_2$)]$^+$ as a charge-transfer adduct.[44] Attachment of the $I_2$ molecule to S(1) affects the Ni–S(1) distances and Ni–S(2)–Ni bond angle. The former lengthen by 0.10 Å and the latter widens by 4.1° relative to **1**. As a consequence, the Ni···Ni distance at 3.601(1) Å in **2** is also longer than in **1** (3.483(1) Å). The angles around the S atoms indicate a pyramidal ($sp^3$) disposition of bonds as in **1** and related complexes with bridging thiophenolate groups.[32] The closest distance between the [Ni$_2$L$^1$($\mu$-OAc)(I$_2$)]$^+$ cation and the pentaiodide is at 4.363(1) Å (I2–I3). This distance is slightly longer than the sum of the van der Waals radii of two iodine atoms indicative of little (if any) secondary bonding interactions between the RS → $I_2$ moiety and the pentaiodide ion. The bond lengths of the $I_5^-$ ion with the I3–I4–I5–I6–I7 sequence (distances of ca. 2.83, 3.02, 3.082, 2.81) are very similar to those observed in other $I_5^-$ structures.[45]

To our knowledge, no diiodine adduct involving a bridging thiophenolate unit has been previously reported. However, a diiodine adduct of an aliphatic nickel thiolate complex NiL'–I$_2$ (H$_2$L' = N,N'-bis(2-mercaptomethylpropane)-1,5-diazacyclooctane) has been communicated by Darensbourg et al.[46] A dinuclear Mo$_2$S$_2$ complex in which each of the bridging sulfido ligands interacts with a $I_2$ molecule has also been reported.[47] The NiL'–I$_2$ complex is characterized by a longer I–I distance (3.016(2) Å) and a shorter S–I distance (2.601(4) Å). It thus appears that short S–I distances are associated with long I–I distances and vice versa,

an observation that has been made previously for many other S–$I_2$ CT adducts.[48-50]

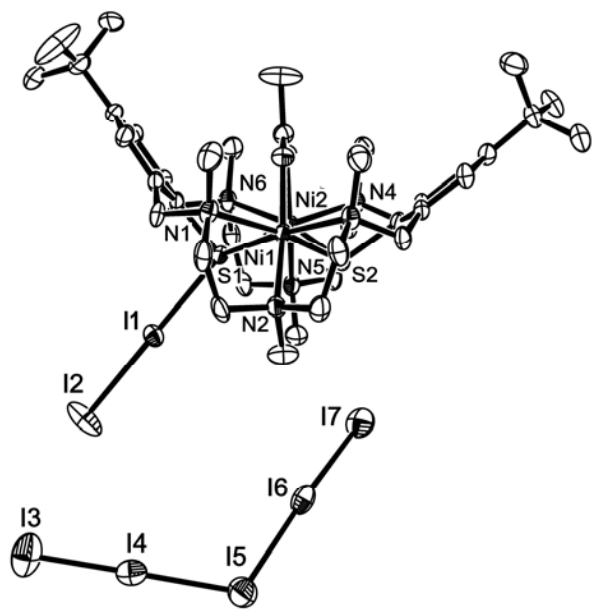

Figure 1. ORTEP representation of the structures of the [Ni$_2$L$^1$($\mu$-OAc)(I$_2$)]$^+$ adduct and the $I_5^-$ anion in crystals of **2**. Thermal ellipsoids are drawn at the 50 % probability level. Hydrogen atoms are omitted for reasons of clarity. Selected bond lengths [Å] and angles [°]: Ni(1)–O(1) 1.988(4), Ni(1)–N(1) 2.285(5), Ni(1)–N(2) 2.138(5), Ni(1)–N(3) 2.188(5), Ni(1)–S(1) 2.593(2), Ni(1)–S(2) 2.449(2), Ni(2)–O(2) 1.986(4), Ni(2)–N(4) 2.191(5), Ni(2)–N(5) 2.148(5), Ni(2)–N(6) 2.301(5), Ni(2)–S(1) 2.572(2), Ni(2)–S(2) 2.446(2), Ni···Ni 3.601(1), Ni(1)–S(1)–N(2) 88.44(5), Ni(1)–S(1)–C(1) 101.8(2), Ni(2)–S(1)–C(1) 101.9(2), Ni(1)–S(1)–I(1) 131.8(2), Ni(2)–S(1)–I(2) 126.5(2), I(1)–S(1)–C(1) 101.6(2) Ni(1)–S(2)–N(2) 94.73(6), Ni(1)–S(2)–C(17) 104.9(2), Ni(2)–S(2)–C(17) 105.6(2); I1–S1 2.755(2), I1–I2 2.894(1), I3–I4 2.832(1), I4–I5 3.009(1), I5–I6 3.082(1), I6–I7 2.813(1), I2···I3 4.363(1), I2···I4 4.624(1); S1–I1–I2 179.54(3), I3–I4–I5 179.32(3), I4–I5–I6 113.45(3), I7–I6–I5 176.73(2).

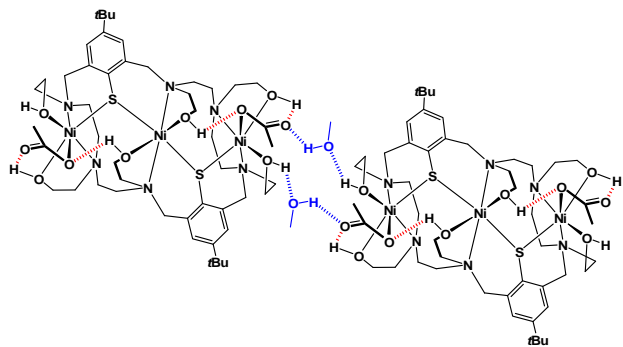

Scheme 3. Inter- and intramolecular hydrogen bonding interactions in solid [Ni$_3$L(OAc)$_2$](BPh$_4$)$_2$ 2MeOH MeCN.

Complex **3**·MeCN·2MeOH crystallizes in the orthorhombic space group P$na2_1$. The structure contains trinuclear [Ni$^{II}_3$L$^2$(OAc)]$^{2+}$ dications, BPh$_4^-$ anions and MeCN and MeOH solvate molecules. The dications and the MeOH solvate molecules are connected by chain hydrogen bonds of the type O$^{Ac}$···HO$^{Me}$···HO involving OH groups of the macrocyclic ligand (O3, O6) and acetate O atoms (O8, O10). The corresponding O···O distances (O8···O12, O12···O3; O10···O11, O11···O6) range from 2.632 to 2.700 Å (see Scheme 3). The other entities are well separated from each other.

Figure 2 presents an ORTEP view of the molecular structure of the trinuclear $[Ni_3L^2(OAc)_2]^{2+}$ dication. Bond lengths and angles are provided in the figure caption. The macrocycle supports a trinuclear complex, with distorted octahedral $N_2O_2S_2$ and $N_2O_3S$ coordination environments for the central and the terminal Ni[II] ions, respectively. This structure is further stabilized by intramolecular OH⋯O hydrogen bonds between the Ni-bound OH groups and the acetate coligands (O1⋯O9 2.776 Å, O2⋯O10 2.591 Å, O4⋯O7 2.790 Å, O5⋯O8 2.568 Å) as indicated in Scheme 3 and Figure 2. The bond lengths and angles around the central and terminal Ni atoms differ slightly, as one might expect. The average Ni–N, Ni–S, and Ni–O distances are 2.098(3) Å, 2.316(1) Å and 2.091(2) Å for the terminal Ni ions and 2.233(3) Å, 2.296(1) Å and 2.132(2) Å for the central one, respectively. The Ni–N and Ni–O distances are normal for six-coordinate Ni[II] complexes. The Ni–S bonds are relatively short. Thus, in dinuclear Ni[II] complexes supported by the parent $H_2L^1$ macrocycle, the Ni–S bonds are much longer and range from 2.45 to 2.52 Å. The angles around the S atoms indicate a pyramidal ($sp^3$) disposition of bonds as in **2**. The Ni–S–Ni bridging angle of 132.8° and 133.5° are, however, much larger than in **2**. The widening of the angles can be attributed to the fact that the Ni ions lie on opposite faces of the thiophenolate planes, and this in turn to the bonding constraints of the macrocycle. In **1** or **2** the Ni atoms are on the same side of the thiophenolate plane, and the Ni–S–Ni angles typically do not deviate much from 90°. It should be noted, however, that Cu–S–Cu angles for $[Cu_2(L-S)(pz)]$ are in the range 99.5(2) – 101.5(2)°,[32] and 128.6(3)° in $[Cu_2(L-S)_2(SC_6H_4Me-p)]^+$.[51] The Ni⋯Ni distances in **3** at 4.229 Å and 4.234 Å are also much longer than in **1** and **2**.

## Magnetic Properties of Complex 2

**Static magnetization:** The field dependence of the static magnetization $M(B)$ of the $Ni_2$-complex **2** at low temperatures is presented in Figure 3. The measurements at $T = 2$ K reveal a saturation magnetization of about 4.4 $\mu_B$/2Ni at $B = 5$ T, which corresponds to the magnetic ground state of the molecule with a total spin $S_{tot} = 2$. This observation implies ferromagnetic (FM) coupling between the two $Ni^{2+}$ ions ($3d^8$, $S_{Ni} = 1$). This conclusion is confirmed by the temperature dependence of the static susceptibility $\chi(T) = M/B$, the inverse susceptibility $\chi^{-1}(T)$ and the product $\chi T$ of the $Ni_2$-complex presented in Figure 4. As will be shown below, the data can be well described by means of the FM dimer model.

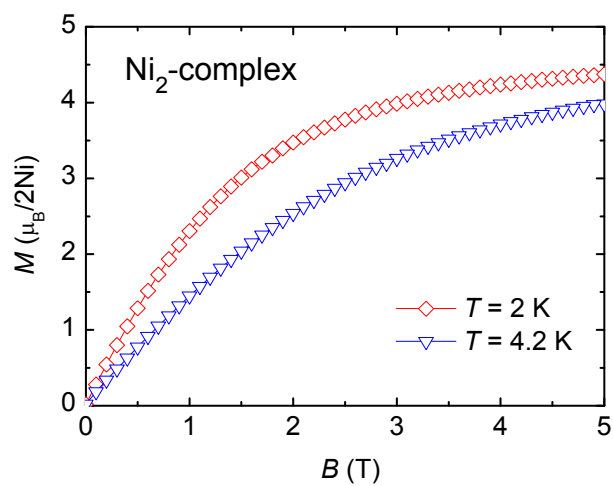

Figure 3. Field dependence of the magnetization $M(B)$ of the $Ni_2$-complex **2** at $T = 2$ K and $T = 4.2$ K.

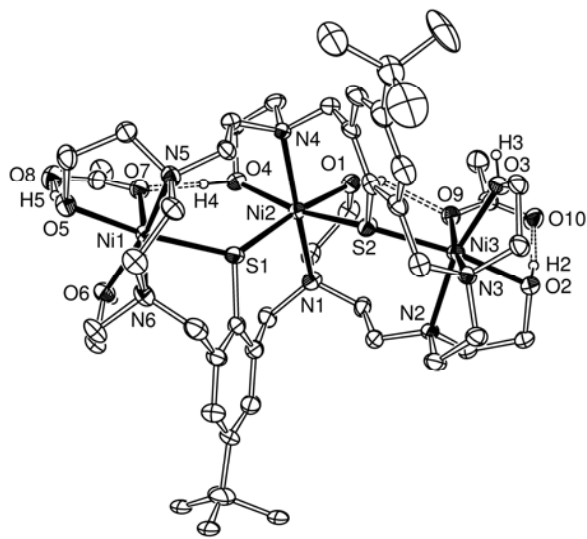

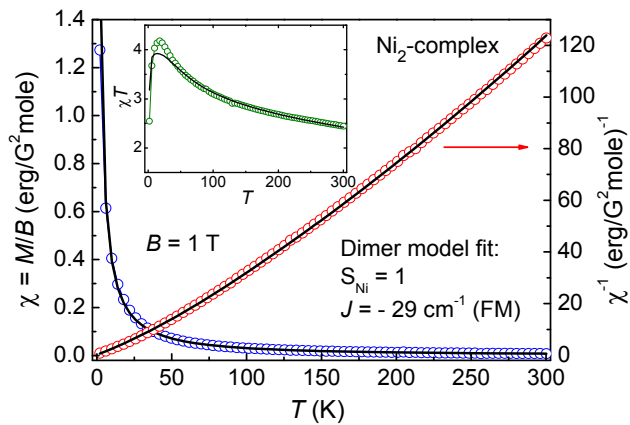

Figure 2. ORTEP representation of the structure of the $[Ni_2L^2(OAc)_2]^{2+}$ dication in crystals of **3**. Hydrogen atoms have been omitted for clarity except for OH groups. The intramolecular hydrogen bonding interactions are shown as dashed lines. Selected bond lengths [Å] and angles [°]: Ni1−N5 2.110(4), Ni1−N6 2.087(3), Ni1−O5 2.127(3), Ni1−O6 2.053(3), Ni1−O7 2.097(3), Ni1−S1 2.318(1), Ni2−O1 2.124(3), Ni2−O4 2.140(3), Ni2−N1 2.246(4), Ni2−N4 2.219(4), Ni2−S1 2.296(1), Ni2−S2 2.295(1), Ni3−O2 2.118(3), Ni3−O3 2.056(3), Ni3−O9 2.096(3), Ni3−N3 2.084(3), Ni3−N2 2.112(3), Ni3−S2 2.313(1); Ni1−S1−Ni2 132.8(1), Ni2−S2−Ni3 133.5(1).

Figure 4. Temperature dependencies of the static susceptibility $\chi(T)$ and the inverse susceptibility $\chi^{-1}(T)$ of the $Ni_2$-complex **2** in $B = 1$ T (open circles). The inset shows the temperature dependence of $\chi T$. The black lines represent a numerical model fit using the Hamiltonian (2 − 4). A temperature independent contribution $\chi_0$ is included into the fit.

**High field ESR:** The HF-ESR of the Ni$_2$-complex was measured on a loose powder sample. The small powder particles were self-oriented in a magnetic field during the measurements due to the presence of the 'easy' axis magnetic anisotropy (see below). A typical ESR spectrum of the Ni$_2$-complex at $T$ = 30 K exhibits four separated ESR lines. The respective frequency $\nu$ vs. resonance field $B_{res}$ dependencies (resonance branches) of all observed lines together with a representative ESR spectrum are shown in Figure 5. The slopes of the resonance branches reveal the $g$-factor of 2.3 for all four ESR lines. The extrapolation of the $\nu(B_{res})$-dependence of line 1 to $B_{res}$ = 0 implies a magnetic anisotropy gap $\Delta \approx$ 70 GHz (2.3 cm$^{-1}$). The $T$-dependence of the ESR spectrum of the Ni$_2$-complex is presented in Figure 6. Here, we observe a shift of the spectral weight to lower magnetic fields at low temperatures, which indicates a negative axial magnetic anisotropy of the molecule ($D_S <$ 0).

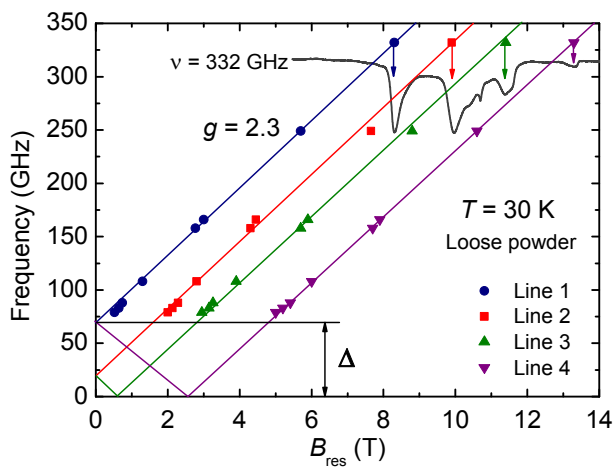

Figure 5. Frequency vs. resonance field $\nu(B_{res})$-dependencies of the ESR resonance modes (resonance branches) for the Ni$_2$-complex **2** at $T$ = 30 K.

**Analysis:** For the analysis of the ESR spectra of the Ni$_2$-complex **2** we introduce a minimal model which describes only the ground state of the molecule. Here, we assume that each molecule has a single total spin $S = S_{tot}$ = 2, and possesses a $g$-factor of 2.3 and a magnetic anisotropy gap $\Delta = |D_S|(S^2 - (S-1)^2)$ = 70 GHz. In this case a minimal effective spin Hamiltonian can be introduced in the form

$$H = D_S(S_z^2 - S(S+1)/3) + E_S(S_x^2 - S_y^2) + g\mu_B \vec{S}\cdot\vec{B} \quad (1)$$

Here, the two first terms describe the zero field splitting of the spin states caused by an anisotropic ligand crystal field potential comprising the axial and the rhombic terms, respectively. Here, $D_S$ is the axial magnetic anisotropy parameter and $E_S$ is the transverse magnetic anisotropy parameter of the complex. Generally $D_S$ and $E_S$ are determined by single ion anisotropies of individual metal ions due to the ligand crystal fields with possible contributions arising from the anisotropic part of the Ni–Ni magnetic exchange interactions. The last term is the Zeeman interaction of the total spin with the external magnetic field $B$. Here $\mu_B$ is the Bohr magneton and $g$ is the $g$-factor. The solution of Hamiltonian (1) yields the eigenvalues $E_i(D,E,B,g)$ and eigenfunctions $\psi_i(D,E,B,g)$ of the energy levels. The knowledge of $E_i$ and $\psi_i$ enables a simulation of the ESR spectra by plugging in the estimates of $D_S$, $E_S$ and $g$ obtained from the raw experimental data. The adjustment of the simulated spectra to the measured ones by fine tuning of $D_S$, $E_S$ and $g$ enables accurate refinement and verification of these parameters.

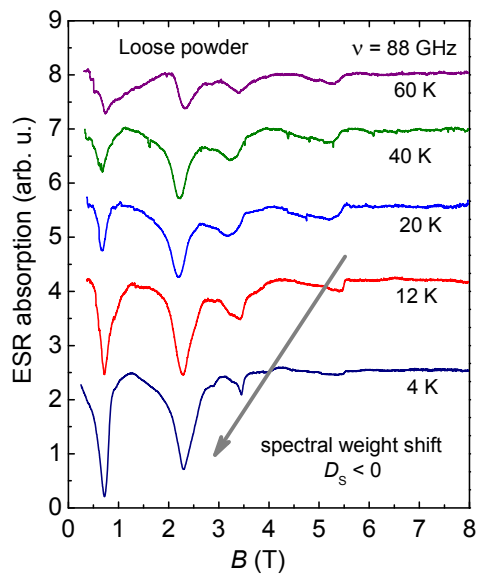

Figure 6. $T$-dependence of the ESR spectrum of the Ni$_2$-complex **2** at $\nu$ = 88 GHz. Note, the relative intensity of the ESR lines changes with the temperature so that the low field lines become dominant.

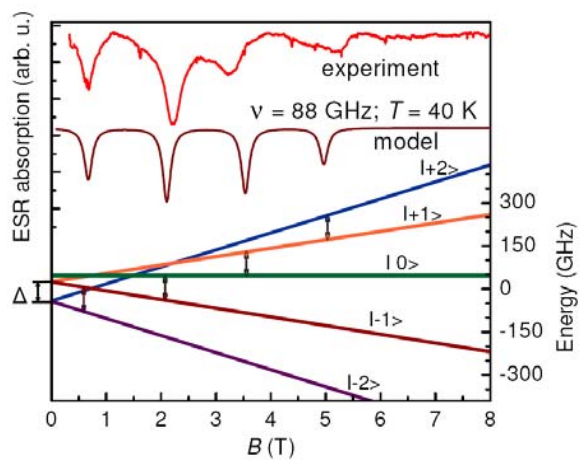

Figure 7. Measured and simulated ESR spectra at $\nu$ = 88 GHz and $T$ = 40 K and the calculated energy level scheme of the spin states of the Ni$_2$-complex.

The simulation of the ESR spectrum was performed for the parallel orientation of the magnetic anisotropy axis to the applied magnetic field. A simulated spectrum at $\nu$ = 88 GHz, $T$ = 40 K together with a representative experimental spectrum of the Ni$_2$-complex and calculated energy levels is presented in Figure 7. The simulation shows that the observed number of the ESR lines (four) is defined by the total spin of the Ni$_2$-complex $S_{tot}$ = 2 and the zero field splitting of the energy levels. Note, that magnetic anisotropy of the Ni$_2$-complex can be well described only by an axial magnetic anisotropy parameter $D_S$ (i.e. $E_S = 0$). The model confirms the experimentally determined value of $\Delta = 3|D_S| = 70$ GHz (2.3 cm$^{-1}$) and the negative sign of the axial magnetic anisotropy $D_S = -23.3$ GHz ($-0.78$ cm$^{-1}$). Note, that the negative sign of $D_S$ implies an easy magnetic anisotropy axis for the Ni$_2$-complex and, therefore, a bistable magnetic ground state.

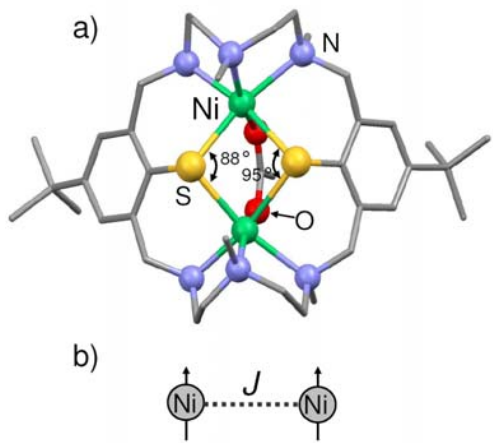

Figure 8. a) Molecular structure of the Ni$_2$-complex **2** (The I$_2$ is omitted for reasons of clarity). b) Scheme of the expected magnetic coupling of the two Ni$^{2+}$ ions in the Ni$_2$-core.

In order to analyze the magnetic properties of the Ni$_2$-complex at higher temperatures, we include in the effective spin Hamiltonian (1) an additional term describing the intramolecular magnetic exchange interactions between individual spins of Ni ions, so that Hamiltonian (1) is upgraded to:

$$H' = \sum_{i>j} J_{ij} \vec{S}_i \cdot \vec{S}_j \quad (2)$$

$$+ D_{eff} \sum_{i=1}^{N} \left(S_{iz}^2 - S_i(S_i+1)/3\right) + E_{eff} \sum_{i=1}^{N} \left(S_{ix}^2 - S_{iy}^2\right) \quad (3)$$

$$+ g\mu_B \sum_{i=1}^{N} \vec{S}_i \cdot \vec{B} \quad (4)$$

Here, (2) describes the isotropic intramolecular magnetic coupling between the neighboring Ni ions ($J < 0$ denotes ferromagnetic coupling). Term (3) describes the zero field splitting of the spin states of each Ni ion and term (4) is the Zeeman interaction of the Ni spins with the external magnetic field $B$. Numerical solution of this Hamiltonian enables a simulation of the temperature dependence of the magnetic susceptibility $\chi(T)$. The topology of the Ni$_2$-core suggests two equivalent exchange paths between the two Ni$^{2+}$ ions effectively described by a single exchange coupling parameter $J$ (cf. Figure 8b). The parameters of the magnetic anisotropy were assumed to be equal for the two Ni ions. Therefore $D_{eff}$ stands here for an effective magnetic anisotropy parameter of each Ni ion which was determined from the ESR data analysis of the spectrum of the ground state spin multiplet $S_{tot} = 2$ as $D_{eff} = D_S(S_{tot}^2 - (S_{tot}-1)^2) = -\Delta$. The simulated $\chi(T)$, $\chi^{-1}(T)$ and $\chi T$ curves are presented together with the experimental data in Figure 4. One can see that the model curves fully reproduce the experimental results. Specifically, the modelling reveals a ferromagnetic coupling $J = -29$ cm$^{-1}$ ($-42$ K) between the two Ni ions and a temperature independent diamagnetic component $\chi_0 = -1.5 \cdot 10^{-3}$ ergG$^{-2}$mole$^{-1}$ caused by the diamagnetic susceptibility of the organic ligands.

## Magnetic Properties of Complex 3

**Static magnetization:** The field dependence of the static magnetization $M(B)$ of the Ni$_3$-complex **3** at 2 K is presented in Figure 9. This measurement reveals a magnetic moment of 1.94 $\mu_B$/3Ni at $B = 5$ T. Although the magnetization is not saturated at $B = 5$ T, the shape of the curve implies an approach to the saturation of the magnetic moment which corresponds to the magnetic ground state of the molecule with a total spin $S^{tot} = 1$. In contrast to the Ni$_2$-complex **2**, $S^{tot} = 1$ here implies an antiferromagnetic (AFM) coupling between the three Ni$^{2+}$ spins ($3d^8$, $S_{Ni} = 1$). The temperature $T$ dependence of the static magnetic susceptibility $\chi = M/B$, the inverse susceptibility $\chi^{-1}$ and the product $\chi T$ value of the Ni$_3$-complex are shown in Figure 10. Here, we observe a strong nonlinearity of $\chi^{-1}(T)$ which, as will be discussed below, corresponds to the thermal activation of higher energy spin multiplets ($S^{tot}_1 = 2$, $S^{tot}_2 = 3$).

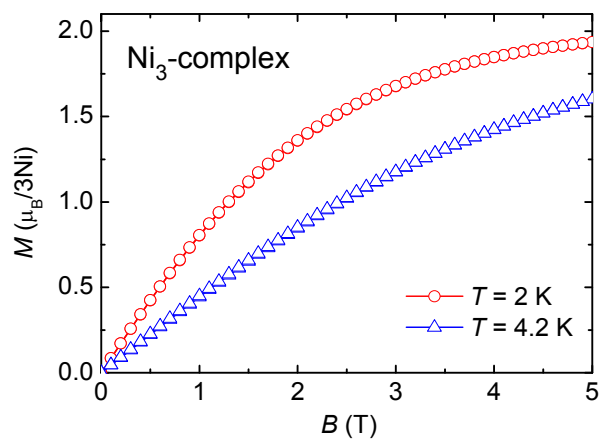

Figure 9. Field dependence of the magnetization $M(B)$ of the Ni$_3$-complex **3** at $T = 2$ K and $T = 4.2$ K.

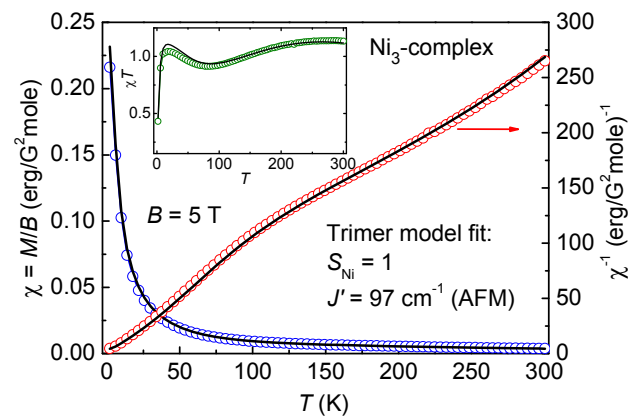

Figure 10. Static susceptibility $\chi(T)$ and inverse susceptibility $\chi^{-1}(T)$ of the Ni$_3$-complex **3** at $B = 5$ T (open circles). The inset shows the temperature dependence of the $\chi T$ value. The black lines represent a numerical model fit using the Hamiltonian (2 – 4). A temperature independent contribution $\chi_0$ is included into the fit.

**High field ESR:** In the case of the Ni$_3$-complex **3**, HF-ESR measurements were performed on a non-oriented powder sample. The fact that the sample was not self-oriented in applied magnetic fields gives the first indication of the easy plane situation for the complex **3**. A typical ESR spectrum of the Ni$_3$-complex at a low temperature ($T$ = 4 K) and the frequency $\nu$ = 166 GHz consists of a sharp stand-alone line at $B_{res}$ ~ 2.5 T (line 1) and a group of lines distributed in the magnetic field range from 3.5 to 7.5 T (lines 2 – 7). The resonance branches of these lines together with a representative ESR spectrum are shown in Figure 11. The slopes of the resonance branches 2 – 7 correspond to the $g$-factor of 2.2. Branch 1 has an almost two times steeper slope which could be described by a phenomenological "effective" $g$-factor of 4.0. Linear extrapolation of $\nu(B_{res})$ of the line 2 to $B_{res}$ = 0 reveals the magnetic anisotropy gap $\Delta \approx$ 60 GHz (2.0 cm$^{-1}$) which, in the case of $S$ = 1, is equal to the axial magnetic anisotropy of the molecule $D_S$.

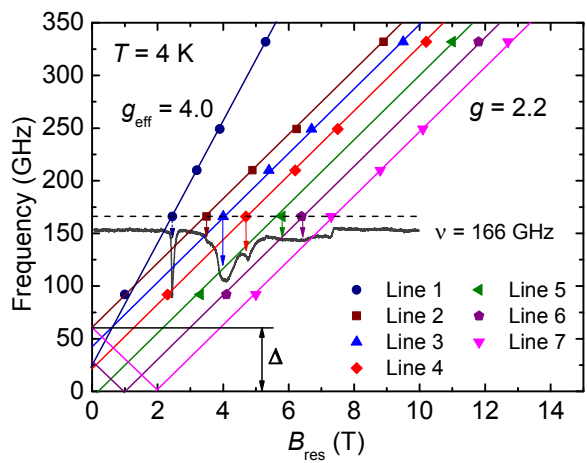

Figure 11. Frequency vs. resonance field $\nu(B_{res})$-dependencies of the ESR resonance modes (resonance branches) for the Ni$_3$-complex **3** at $T$ = 4 K.

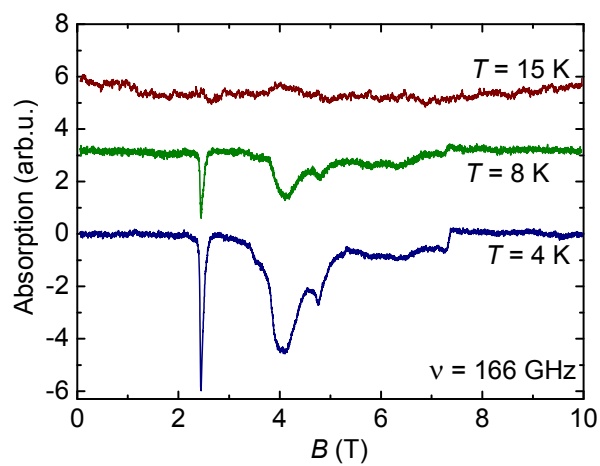

Figure 12. Temperature-dependence of the ESR spectrum of the Ni$_3$-complex **3** at $\nu$ = 166 GHz.

The $T$-dependence of the ESR spectrum is presented in Figure 12. The intensity of the ESR lines decreases with increasing temperature, whereas the relative intensity of the lines does not change. This observation gives evidence that all detected ESR lines correspond to the ground state of the molecule $S^{tot}_0$ = 1. In contrast, the activation of higher energy spin multiplets ($S^{tot}_1$ = 2, $S^{tot}_2$ = 3) is not detected by the ESR measurements, which suggests a large intramolecular coupling rendering the ground state well isolated from higher spin multiplets.

**Analysis:** As it was mentioned before, the ESR spectrum of the Ni$_3$-complex at $T$ = 4 K contains only lines corresponding to the ground state of the molecule ($S^{tot}_0$ = 1). Therefore, for the analysis of the ESR spectrum at $T$ = 4 K we use a minimal effective Hamiltonian (1) which describes only the ground state. Based on our experimental data, we assume that each molecule has a single spin $S = S^{tot}_0$ = 1 with the $g$-factor of 2.2 and the magnetic anisotropy $\Delta = |D_S|(S^2 - (S - 1)^2)$ = 60 GHz. A simulated powder spectrum for $\nu$ = 166 GHz and $T$ = 4 K is presented in Figure 13b. Comparison of the calculated result (b) with the experimental ESR spectrum of a microcrystalline powder sample (a) on that Figure shows that the experimental data are well reproduced by the model. However, we observe a slight difference between the simulated and measured spectra, such as different relative intensities of the lines 3 and 4. We attribute this difference to a partial orientation (texturing) of the microcrystalline powder sample.

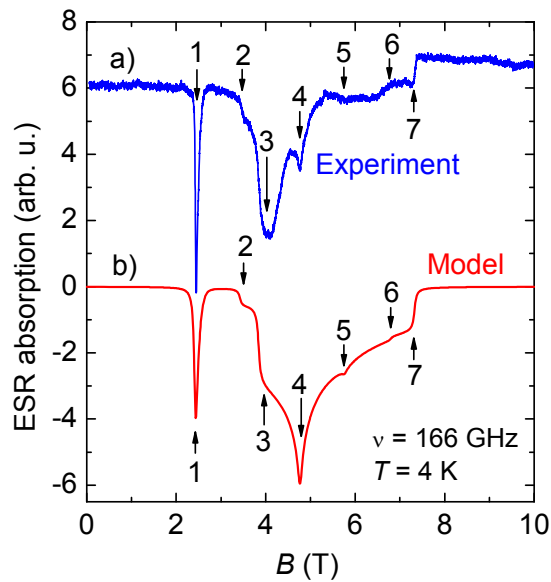

Figure 13. ESR spectra of **3** at $T$ = 4 K and $\nu$ = 166 GHz. a) experimental spectrum; b) simulated spectrum.

Our analysis demonstrates that the applied model describes the Ni$_3$-complex at low temperatures very well. In particular, the ground state of the molecule with $S^{tot}_0$ = 1 and the $g$-factor of 2.2 are confirmed. Moreover, the modelling yields a positive sign of the axial anisotropy $D_S$ = + 60 GHz (2.0 cm$^{-1}$) and a substantial transverse anisotropy $E_S$ = 10 GHz (0.3 cm$^{-1}$) which implies an easy plane situation for the molecule with an easy axis in the plane. In addition, this analysis reveals that the ESR lines 2 – 7 correspond to "allowed" resonance transitions between neighbouring energy levels according to the ESR selection rule $\Delta S_z = \pm 1$ and, therefore, the observed number of lines can be explained by the powder averaging of the ESR spectrum. Line 1 corresponds to the so-called "forbidden" transition with $\Delta S_z = \pm 2$ for which, therefore, an almost doubled slope of the $\nu(B_{res})$ branch is observed ($g_{eff} \sim 2g$, see Figure 11). A non-zero intensity of this transition indicates the mixing of the spin energy states due to the anisotropic ligand crystal field and the spin-orbit coupling.

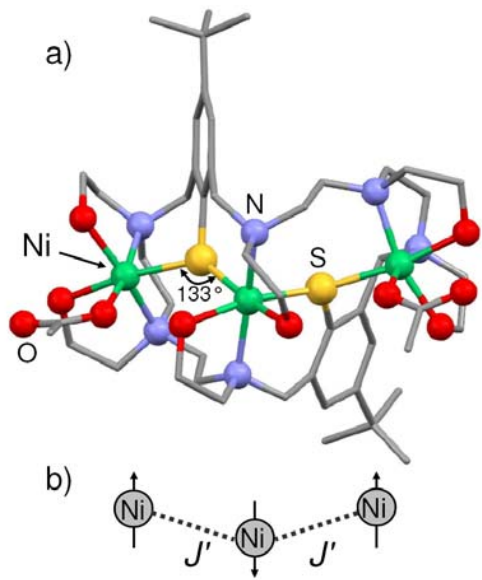

Figure 14. a) Molecular structure of the Ni$_3$-complex **3**. b) Scheme of the expected magnetic coupling of the tree Ni$^{2+}$ ions in the Ni$_3$-core.

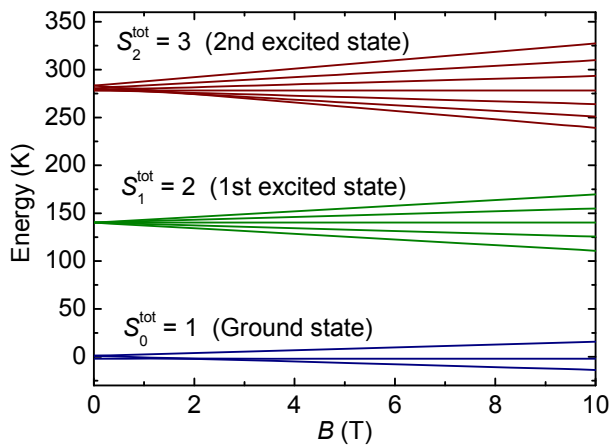

Figure 15. Calculated energy spectrum of the spin states of the Ni$_3$-complex **3**.

The temperature dependence of the magnetic susceptibility of the Ni$_3$-complex can be analyzed by taking into account the topology of the Ni$_3$-core which suggests a single exchange coupling $J'$ between the three Ni$^{2+}$ ions via the sulphur bridge, respectively (cf. Figure 14b). Similar to the Ni$_2$-complex, for the simulation, the parameters of the magnetic anisotropy were assumed to be equal for all three ions and were taken as $D_{eff} = D_S = 2.0$ cm$^{-1}$ and $E_{eff} = E_S = 0.3$ cm$^{-1}$ from the analysis of the ESR spectra. The simulated $\chi(T)$ and $\chi^{-1}(T)$ dependencies fully reproduce the experimental results (cf. the black lines in Figure 10) and reveal an antiferromagnetic coupling $J' = 97$ cm$^{-1}$ (140 K) (AFM) between the Ni ions. According to this model, the energy levels of the spin states of the Ni$_3$-complex have been calculated (see Figure 15). The energy level scheme illustrates the energy gap of 140 K between the ground state $S^{tot}_0 = 1$ and the first excited state $S^{tot}_1 = 2$ caused by the magnetic coupling $J'$. This result confirms that the strong nonlinearity of $\chi^{-1}(T)$ in the temperature range $T = 100 - 150$ K (cf. Figure 10) is caused by the thermal activation of the higher energy spin multiplet ($S^{tot}_1 = 2$). We note, that the modelling reveals a rather large temperature independent diamagnetic component $\chi_0 = -3.9 \cdot 10^{-3}$ ergG$^{-2}$mole$^{-1}$ which is in agreement with a significant diamagnetic susceptibility of the organic ligands of the complex **3**.

## Conclusion

The ability to synthesize polynuclear nickel complexes with quite obtuse Ni-S-Ni angles has been demonstrated with the synthesis of the cationic complexes [Ni$^{II}_2$L$^1$($\mu$-OAc)·I$_2$][I$_5$] (**2**) and [Ni$^{II}_3$L(OAc)$_2$](BPh$_4$)$_2$ (**3**). Both complexes were readily prepared and fully characterized by elemental analysis, IR and UV/vis spectroscopy. We have also investigated the magnetic properties of the novel complexes by means of HF-ESR and static magnetization measurements and analyzed the experimental data by means of minimal spin Hamiltonians.

The structural analysis of the Ni$_2$- and Ni$_3$-complexes revealed a totally different arrangement of the atoms in the cores. In the Ni$_3$-core of **3** adjacent Ni ions are connected via one bridging thiophenolate sulphur atom, whereas in **2** the metal ions are bridged by two $\mu$-S atoms. We observe very different Ni–S–Ni bridging angles in these two complexes. The large value of about 133° in the case of the Ni$_3$-complex is attributed to the fact that the Ni ions lie on opposite faces of the thiophenolate planes. Dinuclear Ni complex **2**, on the other hand, has the Ni ions on the same side of the thiophenolate plane, and shows much smaller Ni–S–Ni angles. The values do not deviate much from 90°.

The HF-ESR and magnetization data show different magnetic exchange interactions in these compounds. The Ni$_2$-complex reveals a ferromagnetic coupling $J = -29$ cm$^{-1}$ (- 42 K), while for the Ni$_3$-complex we observe a strong antiferromagnetic coupling $J' = 97$ cm$^{-1}$ (140 K). These differences may be attributed to the different Ni–S–Ni bridging bond angles, although other effects such as the mutual orientations of the magnetic orbitals (which are different in the two compounds) cannot be ruled out at the moment. A paramount dependence of the magnetic exchange interactions would be in agreement with similar observations made for related binuclear transition metal thiophenolate complexes.[32,52,53] Moreover, the different structure of the ligands in the Ni$_3$- and Ni$_2$-complexes causes opposite signs of the magnetic anisotropy. Thus, in the Ni$_3$-complex the ligand structure yields the positive axial anisotropy and, therefore, an easy plane situation for the molecule. In contrast, for the Ni$_2$-complex we observe the negative axial anisotropy, which yields an easy magnetic anisotropy axis for the molecule and a bistable magnetic ground state.

## Experimental Section

**General:** Unless otherwise noted the preparations were carried out under an argon atmosphere by using standard Schlenk techniques. Complex [Ni$_2$L$^1$(OAc)]ClO$_4$ (**1**·ClO$_4$)[38] and ligand H$_2$L$^{2,[34]}$ were prepared according to literature procedures. All other reagents were obtained from standard commercial sources and used without further purifications. Melting points were determined in capillaries and are uncorrected. The infrared spectra were recorded as KBr discs using a Bruker VECTOR 22 FT-IR-spectrophotometer. Electronic absorption spectra were recorded on a Jasco V-670 UV/vis/near IR spectrophotometer. ESI-FTICR mass spectra were recorded with a Bruker-Daltronics Apex II instrument using dilute MeOH or MeCN solutions.

**Nickel Complex [Ni$_3$(L)(O$_2$CCH$_3$)$_2$(I$_2$)]·I$_5$ (2):** To a solution of [Ni$_2$L$^1$($\mu$-O$_2$CCH$_3$)][ClO$_4$] (93 mg, 0.10 mmol) in acetonitrile (10 mL) was added a solution of I$_2$ (127 mg, 0.50 mmol) in acetonitrile (5 mL), resulting in an immediate color change from pale-green to brown. After standing for 12 h at 0 °C, black crystals formed, which were filtered off, and dried in air. Yield: 125.8 mg, 73 %. M.p. 295 °C (decomp.). IR (KBr): $\nu$ / cm$^{-1}$ = 2957 s, 2858 s (C–H), 1579 s [$\nu_{as}$(CH$_3$CO$_2^-$)], 1484 m, 1457 s, 1432 s [$\nu_s$(CH$_3$CO$_2^-$)], 1361 m, 1307 m, 1264 m, 1235 m, 1198 m, 1154 m, 1073 s, 1037 s, 952 m. UV/VIS (CH$_3$CN): $\lambda_{max}$ / nm ($\varepsilon$ / m$^{-1}$cm$^{-1}$) = 1197 (125), 782 sh (311), 362 (60230), 292 (107514), 204 (137569). Elemental analysis calcd (%) for C$_{40}$H$_{67}$I$_7$N$_6$Ni$_2$O$_2$S$_2$ (1733.85): C 27.71, H 3.89, N 4.85; found: C 27.10, H 3.72, N 4.59. This compound was additionally characterized by X-ray crystal structure analysis.

**Nickel Complex [Ni$_3$L$^2$(O$_2$CCH$_3$)$_2$](BPh$_4$)$_2$ (3):** To a suspension of H$_2$L$^2$·6HCl (952 mg, 0.890 mmol) in methanol (20 mL) was added under nitrogen a solution of Ni(O$_2$CCH$_3$)$_2$·4H$_2$O (507 mg, 1.78 mmol) in methanol (2 mL), followed by a solution of triethylamine (719 mg, 7.12 mmol) in methanol (1 mL). The resulting deep green solution was stirred for further 12 h, a small amount of an insoluble material was filtered off, and to the clear filtrate was added solid NaBPh$_4$ (1.53 g, 4.47 mmol). The resulting solid was filtered, washed with MeOH (2 × 20 mL). Green diamond-shaped crystals of the title compound (1.02 g, 64 %) formed upon recrystallization from acetonitrile. IR (KBr): $\nu$ / cm$^{-1}$ = 3422 m, 3200 sh, 3054 s, 2965 s, 2902 m, 2867 m, 2360 w, 1944 vw, 1826 vw, 1771 vw, 1614 w, 1579 m, 1541 m, 1478 m, 1459 s, 1426 s, 1402 m, 1371 m, 1337 m, 1310 m, 1244 s, 1183 m, 1153 m, 1129 m, 612 m, 1091 m, 1059 s, 1033 m, 985 m, 953 w, 896 m, 862 w, 846 w, 799 w, 733 [s, v(BPh$_4^-$)], 706 [vs, v(BPh$_4^-$)], 663 w. UV/VIS (CH$_3$CN): $\lambda_{max}$ / nm ($\varepsilon$ / m$^{-1}$cm$^{-1}$) = 644 (83), 903sh (53), 1073 (95). MS (ESI, +ve, CH$_3$CN): m/z = 1459.5 (M-BPh$_4$)$^+$. Elemental analysis calcd (%) for C$_{96}$H$_{122}$B$_2$N$_6$Ni$_3$O$_{10}$S$_2$ (1781.86): C 64.71, H 6.90, N 4.72, S 3.60; found: C 64.30, H 6.78, N 4.55, S 3.41. This compound was additionally characterized by X-ray crystal structure analysis.

**Crystal structure determination:** A single crystal of **2** suitable for X-ray structure analysis was picked from the reaction mixture. Crystals of **3**·MeCN·2MeOH were grown by slow evaporation of a mixed acetonitrile/methanol (1:1) solution. The data sets were collected at 213(2) K using a STOE IPDS-2T diffractometer and graphite monochromated Mo-K$\alpha$ radiation (0.71073 Å). The intensity data were processed with the program STOE X-AREA. The structure was solved by direct methods[54] and refined by full-matrix least-squares on the basis of all data against $F^2$ using SHELXL-97.[55] PLATON was used to search for higher symmetry.[56] Drawings were produced with Ortep-3 for Windows.[57] Unless otherwise specified the H atoms were placed at calculated positions and refined as riding atoms with isotropic displacement parameters. All non-hydrogen atoms were refined anisotropically.

*Crystal data for **2***: C$_{40}$H$_{67}$I$_7$N$_6$Ni$_2$O$_2$S$_2$, $M_r$ = 1733.84, triclinic, space group $P$-1, $a$ = 14.222(3) Å, $b$ = 15.058(3) Å, $c$ = 15.703(3) Å, $\alpha$ = 69.28(3)°, $\beta$ = 64.97(3)°, $\gamma$ = 70.16(3)°, $V$ = 2778(1) Å$^3$, $Z$ = 2, $\rho_{calcd}$ = 2.073 g cm$^{-3}$; $T$ = - 160 °C, $\mu$(Mo$_{K\alpha}$) 4.680 mm$^{-1}$ ($\lambda$ = 0.71073 Å); 22298 reflections measured, 10217 unique, 6969 with $I > 2\sigma(I)$, refinement converged to $R$ = 0.0403, $wR$ = 0.0915 ($I >2\sigma(I)$), 532 parameters and 0 restraints, min./max. residual electron density = +1.858/ –1.543 e/Å$^3$.

*Crystal data for **3**·MeCN·2MeOH*: C$_{100}$H$_{133}$B$_2$N$_7$Ni$_3$O$_{12}$S$_2$, $M_r$ = 1887.00, orthorhombic, space group P$na$2$_1$, $a$ = 26.464(5) Å, $b$ = 16.312(3) Å, $c$ = 22.480(5) Å, $V$ = 9704(3) Å$^3$, $Z$ = 4, $\rho_{calcd}$ = 1.292 g cm$^{-3}$; $\mu$ (Mo$_{K\alpha}$) = 0.682 mm$^{-1}$ ($\lambda$ = 0.71073 Å); 118807 reflections measured, 24527 unique, 13700 with $I >2\sigma(I)$, refinement converged to $R$ = 0.0456, $wR$ = 0.0981 ($I > 2\sigma(I)$), 1167 parameters. The Flack $x$ parameter (absolute structure parameter) for **3** was calculated to be 0.45(1) indicative of inversion twinning. This inversion twinning was modeled by using the TWIN and BASF instructions implemented in the SHELXTL software package, to give a BASF factor of 0.50. In addition, a split atom model was used to account for the disorder of one $t$Bu group resulting in site occupancies of 0.38(1) for C26a-C28a and 0.62(1) for C26b-C28b, respectively. The six OH hydrogen atoms were located from final difference Fourier maps but their positions were refined as riding atoms with isotropic displacement parameters, min./max. residual electron density = +0.367/ –0.643 e/Å$^3$.

CCDC-755183 (**2**) and 755184 (**3**) contain the supplementary crystallographic data for this paper. These data can be obtained free of charge from The Cambridge Crystallographic Data Centre via www.ccdc.cam.ac.uk/data_request/cif.

**Magnetic measurements:** The static magnetization $M$ and the high field electron spin resonance (HF-ESR) of the Ni-complexes were studied on microcrystalline powder samples. Magnetic field $B$ and temperature $T$ dependencies of the static magnetization were measured in the temperature range $T$ = 2 – 300 K and in magnetic fields $B$ up to 5 T with a commercial SQUID (Superconducting Quantum Interference Device) magnetometer MPMS-XL5 (Quantum Design). HF-ESR in static magnetic fields was studied with a Millimeterwave Vector Network Analyzer (AB Millimétré) used for generation of millimeter- and submillimeter microwaves and phase locked detection of a signal.[35] The measurements have been performed in the frequency range from 80 GHz up to 350 GHz and in magnetic fields up to 15 T. The analysis and the simulation of the ESR spectra were done by means of the EasySpin toolbox for Matlab.[58] The analysis and the simulation of the temperature dependence of the magnetic susceptibility $\chi(T)$ were done by means of the *julX* simulation program.[59]

## *Acknowledgements*

*We are particularly grateful to Prof. Dr. H. Krautscheid for providing facilities for X-ray crystallographic measurements. Financial support of this work from the Deutsche Forschungsgemeinschaft (Priority Program "Molecular Magnetism") and by the Universität Leipzig is gratefully acknowledged. A.A. and A.P. gratefully acknowledge funding by the International Max Planck Research School for Dynamical Processes in Atoms, Molecules and Solids, Y.K. and R.K. have been supported by the DFG via KL 1824/2.*

**Keywords:** ((MAGN. PROP. · ESR · Ni · Compl. · Crystal-struct. determination))

[1] I.G. Dance, *Polyhedron* **1986**, *5*, 1037–1104.
[2] B. Krebs, G. Henkel, *Angew. Chem.* **1991**, *105*, 785–804; *Angew. Chem. Int. Ed. Engl.* **1991**, *30*, 769–788.
[3] M.A. Halcrow, G. Christou, *Chem. Rev.* **1994**, *94*, 2421–2481.
[4] P.J. Blower, J.R. Dilworth, *Coord. Chem. Rev.* **1987**, *76*, 121–185.
[5] G. Henkel, B. Krebs, *Chem. Rev.* **2004**, *104*, 801–824.
[6] C.A. Grapperhaus, M.J. Darensbourg, *Acc. Chem. Res.* **1998**, *31*, 451–459.
[7] D. Sellmann, J. Sutter, *Acc. Chem. Res.* **1997**, *30*, 460–469.

[8] A.C. Marr, D.J.E. Spencer, M. Schröder, *Coord. Chem. Rev.* **2001**, *219–221*, 1055–1074.
[9] T.C. Harrop, P.K. Mascharak, in *Concepts and Models in Bioinorganic Chemistry*, H.-B. Kraatz, N. Metzler-Nolte, 1st ed., Wiley-VCh, **2006**, p. 309–329.
[10] T. Beissel, K.S. Bürger, G. Voigt, K. Wieghardt, C. Butzlaff, A.X. Trautwein, *Inorg. Chem.* **1993**, *32*, 124–126.
[11] T. Glaser, T. Beissel, E. Bill, T. Weyhermüller, V. Schünemann, W. Meyer-Klaucke, A.X. Trautwein, K. Wieghardt, *J. Am. Chem. Soc.* **1999**, *121*, 2193–2208.
[12] H.-J. Krüger, R.H. Holm, *J. Am. Chem. Soc.* **1990**, *112*, 2955–2963.
[13] P. Zanello, S. Tamburini, P.A. Vigato, G.A. Mazzochin, *Coord. Chem. Rev.* **1987**, *77*, 165–273.
[14] S. Brooker, *Coord. Chem. Rev.* **2001**, *222*, 33–56.
[15] A.J. Atkins, D. Black, A.J. Blake, A. Marin-Becerra, S. Parsons, L. Ruiz-Ramirez, M. Schröder, *Chem. Commun.* **1996**, 457–464.
[16] C. Loose, V. Lozan, J. Kortus, B. Kersting, *Coord. Chem. Rev.* **2009**, *253*, 2244–2260.
[17] B. Kersting, U. Lehmann, in *Adv. Inorg. Chem.* **2009**, *61*, 407–470, (ed. R. v. Eldik, C. D. Hubbard).
[18] J. Hausmann, M.H. Klingele, V. Lozan, G. Steinfeld, D. Siebert, Y. Journaux, J.J. Girerd, B. Kersting, *Chem. Eur. J.* **2004**, *10*, 1716–1728.
[19] Y. Journaux, J. Hausmann, V. Lozan, B. Kersting, *Chem. Commun.* **2006**, 83–84.
[20] T. Glaser, Y. Journaux, G. Steinfeld, V. Lozan, B. Kersting, *J. Chem. Soc. Dalton Trans.* **2006**, 1738–1748.
[21] V. Lozan, B. Kersting, *Inorg. Chem.* **2008**, *47*, 5386–5393.
[22] J. Goodenough, Magnetism and the Chemical Bond. John Wiley and Sons, New York, **1963**.
[23] V.Yu. Yushankhai, R. Hayn, *Europhys. Lett.* **1999**, *47*, 116–121.
[24] S. Tornow, O. Entin-Wohlman, A. Aharony, *Phys. Rev. B* **1999**, *60*, 10206–10215.
[25] V. Kataev, K.-Y. Choi, M. Grüninger, U. Ammerahl, B. Büchner, A. Freimuth, A. Revcolevschi, *Phys. Rev. Lett.* **2001**, *86*, 2882–2885.
[26] W.E.A. Lorenz, R.O. Kuzian, S.-L. Drechsler, W.-D. Stein, N. Wizent, G. Behr, J. Malek, U. Nitzsche, H. Rosner, A. Hiess, W. Schmidt, R. Klingeler, M. Loewenhaupt, B. Büchner, *EPL* **2009**, *88*, 37002.
[27] W. Geertsma, D. Khomskii, *Phys. Rev. B* **1996**, *54*, 3011–2014.
[28] O. Kahn, Molecular Magnetism, Wiley-VCH, Weinheim, **1993**.
[29] Magnetism: Molecules to Materials, ed. J.S. Miller, M. Drillon, Wiley-VCH, Weinheim, **2001**.
[30] E. Coronado, P. Delhaes, D. Gatteschi, J.S. Miller, Eds., Molecular magnetism: From Molecular Assemblies to Devices, NATO ASI Series, Kluwer, Dordrecht: The Netherlands, **1995**, Vol. 321.
[31] (a) T. Beissel, F. Birkelbach, E. Bill, T. Glaser, F. Kesting, C. Krebs, T. Weyhermüller, K. Wieghardt, C. Butzlaff, A.X. Trautwein, *J. Am. Chem. Soc.* **1996**, *118*, 12376–12390. (b) U. Bossek, D. Nühlen, E. Bill, T. Glaser, C. Krebs, T. Weyhermüller, K. Wieghardt, M. Lengen, A.X. Trautwein, *Inorg. Chem.* **1997**, *36*, 2834–2843.
[32] (a) P. Iliopoulos, K.S. Murray, R. Robson, J. Wilson, G.A. Williams, *J. Chem. Soc. Dalton Trans.* **1987**, 1585–1591. (b) P. Iliopoulos, G.D. Fallon, K.S. Murray, *J. Chem. Soc. Dalton Trans.* **1988**, 1823–1826.
[33] G. Steinfeld, V. Lozan, H.-J. Krüger, B. Kersting, *Angew. Chem.* **2009**, *121*, 1988–1991; *Angew. Chem. Int. Ed.* **2009**, *48*, 1954–1957.
[34] M. Gressenbuch, B. Kersting, *Dalton Trans.* **2009**, 5281–5283.
[35] C. Golze, A. Alfonsov, R. Klingeler, B. Büchner, V. Kataev, C. Mennerich, H.-H. Klauss, M. Goiran, J.-M. Broto, H. Rakoto, S. Demeshko, G. Leibeling, F. Meyer, *Phys. Rev. B* **2006**, *73*, 224403.
[36] P. Chaudhuri, V. Kataev, B. Büchner, H.-H. Klauss, B. Kersting, F. Meyer, *Coord. Chem. Rev.* **2009**, *253*, 2261–2285.
[37] C. Mennerich, H.-H. Klauss, M. Broekelmann, F.J. Litterst, C. Golze, R. Klingeler, V. Kataev, B. Büchner, S.-N. Grossjohann, W. Brenig, M. Goiran, H. Rakoto, J.-M. Broto, O. Kataeva, D.-J. Price, *Phys. Rev. B* **2006**, *73*, 174415.
[38] B. Kersting, *Angew. Chem.* **2001**, *113*, 4110–4112; *Angew. Chem. Int. Ed.* **2001**, *40*, 3988–3990.
[39] K. Nakamoto, *Infrared and Raman Spectra of Inorganic and Coordination Compounds*, Wiley, New York, **1978**.
[40] N.A. Al-Hashimi, K.A. Hassan, E-M. Nour, *Spectrochmica Acta* **2005**, *62A*, 317–321.
[41] W. Gabes, D.J. Stufkens, *Spectrochimica Acta* **1974**, *30A*, 1835–1841.
[42] D.J. Seery, D. Britton, *J. Phys. Chem.* **1964**, *68*, 2263–2266.
[43] A. B. P. Lever, *Inorganic Electronic Spectroscopy*, 2nd edition, Elsevier Science, Amsterdam, **1984**.
[44] I. L. Karle, *J. Chem. Phys.* **1955**, 23, 1739–1740.
[45] C. Horn, M. Scudder, I. Dance, *CrystEngComm.* **2001**, 1–6.
[46] E.J. Lyon, G. Musi, J.H. Reibenspies, M.H. Darensbourg, *Inorg. Chem.* **1998**, *37*, 6942–6949.
[47] J. Allshouse, R.C. Haltiwanger, V. Allured, M.R. DuBois, *Inorg. Chem.* **1994**, *33*, 2505–2506.
[48] F.H. Herbstein, W. Schwotzer, *J. Am. Chem. Soc.* **1984**, *106*, 2367–2373.
[49] F. Bigoli, P. Deplano, M.L. Mercuri, M.A. Pellinghelli, A. Sabatini, E.F. Trogu, A. Vacca, *J. Chem. Soc. Dalton Trans.* **1996**, 3583–3589.
[50] A.J. Blake, F.A. Devillanova, A. Garau, L.M. Gilby, R.O. Gould, F. Isaia, F. Isaia, V. Lippolis, S. Parsons, C. Radek, M. Schröder, *J. Chem. Soc. Dalton Trans.* **1998**, 2037–2046.
[51] N. Aoi, G. Matsubayashi, T. Tanaka, *J. Chem. Soc. Dalton Trans.* **1987**, 241–247
[52] J.R. Dorfman, J.J. Girerd, E.D. Simhon, T.D.P. Stack, R.H. Holm, *Inorg. Chem.* **1984**, *23*, 4407–4416.
[53] P.J. Hay, J.C. Thibeault, R. Hoffmann, *J. Am. Chem. Soc.* **1975**, *97*, 4884–4889.
[54] G.M. Sheldrick, *Acta Crystallogr.* **1990**, *A46*, 467–473.
[55] G.M. Sheldrick, SHELXL-97, Computer program for crystal structure refinement, University of Göttingen, Göttingen, Germany, **1997**.
[56] A.L. Spek, PLATON - A Multipurpose Crystallographic Tool; Utrecht University, Utrecht, The Netherlands, **2000**.
[57] L.J. Farrugia, *J. Appl. Crystallogr.* **1997**, *30*, 565–568.
[58] S. Stoll, A. Schweiger, *J. Magn. Reson.* **2006**, *178*(1), 42–55.
[59] http://ewww-mpi-muelheim.mpg.de/bac/logins/bill/julX en.php